\documentclass[aps,pra,twocolumn,longbibliography,superscriptaddress]{revtex4-2}

\usepackage{dcolumn}
\usepackage{bm}
\usepackage{subfigure}
\usepackage[normalem]{ulem}
\usepackage{enumerate}
\usepackage{cancel}
\usepackage{flushend}
\usepackage{mathtools}
\usepackage{float}
\usepackage{stmaryrd}
\usepackage{colortbl}
\usepackage[table]{xcolor}
\usepackage{array,ragged2e}
\usepackage{amssymb}
\usepackage{wasysym,graphicx,multirow,textcomp}
\usepackage{url}
\usepackage{romannum}

\usepackage[colorlinks = true,linkcolor = red,citecolor = blue]{hyperref}
\usepackage[sort&compress]{natbib}

\usepackage{tikz-cd}

\newcommand{\deri}{\text{d}}

\newcommand{\eq}[1]{Eq.~(\ref{#1})}

\newcommand{\quoting}[1]{``#1''}
\newcommand{\fig}[1]{Fig.~\ref{#1}}
\newcommand{\h}{h_\text{eff}}

\usepackage{lineno}
\begin{document}

\title{Decay Rates of Optical Modes Unveil the Island Structures in Mixed Phase Space}

    \author {Chang-Hwan Yi}
%   \email{yichanghwan@ibs.re.kr}
    \affiliation{Center for Theoretical Physics of Complex Systems, IBS, Daejeon 34126, Korea}
    \author {Barbara Dietz}
%   \email{bdietzp@gmail.com}    
    
    \affiliation{Center for Theoretical Physics of Complex Systems, IBS, Daejeon 34126, Korea}
 \affiliation{Basic Science Program, Korea University of Science and Technology (UST), Daejeon 34113, Korea}
    \author {Jae-Ho Han}
%    \email{jaehohan@kaist.com}
    \thanks{Corresponding author: jaehohan@kaist.ac.kr}
    \affiliation{Department of Physics, Korea Advanced Institute of Science and Technology (KAIST), Daejeon 34141, Korea}
    
    \author {Jung-Wan Ryu}
%    \email{jungwanryu@gmail.com}
    \thanks{Corresponding author: jungwanryu@gmail.com}
    \affiliation{Center for Theoretical Physics of Complex Systems, IBS, Daejeon 34126, Korea}
    \affiliation{Basic Science Program, Korea University of Science and Technology (UST), Daejeon 34113, Korea}

\begin{abstract}
We explore the decay rates of optical modes in asymmetric microcavities with mixed phase space across a wide range of wavelengths that extend deep into the semiclassical, i.e., short-wavelength limit. Implementing an efficient numerical method, we computed $10^6$ eigenmodes and discovered that certain decay rates form sequential separate branches with increasing wavenumber that eventually merge into smooth curves. The analysis of the localization properties and Husimi distributions reveals that each branch corresponds to a periodic orbit in the closed classical system. Our findings show that these decay rates gradually resolve the structure of the islands in mixed phase space as we approach the short-wavelength limit. We present an effective semiclassical model incorporating wavenumber-dependent localization, Fresnel reflection, and the Goos-H\"anchen shift and demonstrate that these effects are crucial in accounting for the observed branches of decay rate curves.
\end{abstract}

\pacs{05.45.Mt, 42.60.Da, 05.45.-a}%, 42.65.Sf, 05.10..a, 05.65.+b}

%\setboolean{displaycopyright}{true}

\maketitle

\section{Introduction\label{Intro}}
Asymmetric optical microcavities (AOMs) have emerged as an active research topic due to their rich physical properties~\cite{RevModPhys.87.61}, which stimulate academic interest and hold potential for various applications, such as high-Q directional light sources~\cite{PhysRevLett.100.033901,PhysRevLett.105.103902,PhysRevLett.104.163902,PhysRevLett.108.243902,ffchaos1,ffchaos2,yi2009lasing} for optoelectronic circuits and ultra-fine-resolution sensors~\cite{wiersig2014enhancing,sensor0,sensor1,sensor2,sensor3,hodaei2017enhanced} in the presence of an exceptional point~\cite{ep1,ep2,ep3,ep4,ep5,ep6,ep7,ep10,ep12,ep16}. The primary research objective of AOMs lies in quantum and wave chaos~\cite{quantumchaos1,LesHouches1989,quantumchaos0} and semiclassical (short-wavelength) approaches, which seek signatures of classical chaos in the properties of the associated quantum or wave-dynamical system~\cite{stone_r_w,LSAYI,PhysRevLett.116.203903}. More precisely, the study of AOMs explores the intricate link between mixed integrable-chaotic classical ray dynamics and non-Hermitian quantum and wave dynamics realized by partially opening the system, thereby providing fundamental insights into the physics of open quantum and wave systems~\cite{PhysRevE.94.022202,PhysRevE.100.042219,ep14,ep9,ep13,ep8,ep11,ep15,PhysRevLett.129.193901}.

The theoretical studies of two-dimensional AOMs are governed by the scalar Helmholtz equation, which is mathematically equivalent to the Schr\"odinger equation of the corresponding quantum billiard (QB)~\cite{Sinai1970,Bunimovich1979} with appropriate boundary conditions~\cite{PhysRevLett.104.163902,stone_bill,harayama_bill0,harayama_bill1}. Accordingly, methods developed for the latter, within the context of quantum chaos, can be used to analyze the properties of AOMs. A classical circular billiard~\footnote{With 'classical billiard' or 'billiard,' we refer to a billiard with hard walls} has integrable dynamics, implying that its phase space comprises only invariant tori. The eigenstates of a circular QB are known analytically, and those of a circular microcavity are well-defined by pairs of quantum numbers, one of which is related to angular momentum in the semiclassical regime (SR) of large wavenumbers through the Einstein-Brillouin-Keller (EBK) quantization~\cite{Keller1960,JUNockel}. Based on EBK, a trace formula has been derived for circular QBs~\cite{Berry1977} and for dielectric cavities~\cite{PhysRevE.78.056202}, which has been tested experimentally in~\cite{PhysRevE.81.066215,PhysRevE.83.036208} and for dielectric cavities shaped like chaotic billiards~\cite{PhysRevE.85.056203}. We consider AOMs resulting from the deformation of a circular cavity, whose dynamics is mixed regular-chaotic, implying that the two-dimensional classical Poincar\'e surface of section (PSOS) exhibits stable resonance chains and chaotic layers, making it infeasible to assign exact angular-momentum modal numbers to eigenmodes. Revealing the correspondence between classical phase space and optical modes in such AOMs remains challenging.

In this context, a fascinating yet not well-explored topic~\cite{Schomerus2004,Schomerus2009,Ketzmerick2025} is the convergence properties of the decay rates of modes in AOMs as the wavenumber $k$ increases deep into the SR. Unlike Hermitian systems with real-valued eigenvalues, such as QBs with Dirichlet boundary conditions, an AOM is a non-Hermitian system with complex-valued eigenvalues. Their imaginary parts determine the decay rates, whose behavior changes as the system transitions from the wave-dynamical regime to deep into the SR regime. We demonstrate that they gradually resolve the classical phase space structure of the corresponding billiard as we enter the SR and thus provide an effective measure for exploring this transition region. In~\cite{PhysRevA.77.013804}, an approximation for the decay rates of modes in dielectric cavities with a slightly deformed circular shape with radius $R$ was derived using a perturbative expansion and verified in~\cite{PhysRevE.100.042219}. By contrast, we obtain optical modes employing a direct numerical method in an AOM with refractive index $n$, yielding over $10^6$ eigenmodes with effective wavenumbers reaching $nkR\simeq2000$. This value is 4–10 times higher than in previous studies and lies well beyond the range where the perturbative approximation is applicable.

Although it is well understood that modes in the extreme short-wavelength limit eventually follow ray optics, the precise nature of the transition regime between wave and ray optics remains an ongoing research topic. In this context, developing an optimal numerical method to obtain accurate solutions in the complex plane at minimized computational cost is a crucial yet challenging issue. With this achievement, we can explore the characteristics of decay rates Im$(nkR)$ at unprecedentedly high wavenumbers. This led to the discovery that the decay rates are grouped into branches that merge with increasing wavenumber into nearly Re$(nkR)$-independent curves. Moreover, each of these curves corresponds to a mode localized within a different stable resonance chain in phase space. This consistent behavior confirms that the decay rates reflect the island structure observed in the PSOS.
        \begin{figure}[t]
        \centering
        \vspace{-0.5cm}
        \includegraphics[width=0.8\columnwidth]{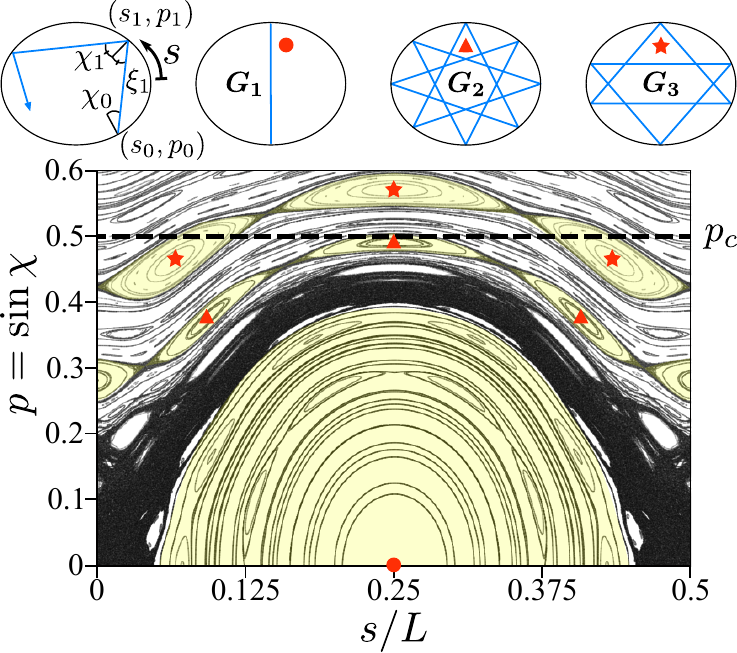}
        \vspace{-0.4cm}
        \caption{Upper half of the PSOS of the quadrupole billiard [\eq{bd}] for $\varepsilon=0.05$, which has mirror symmetries with respect to the two diameter axes. The yellow-shaded regions highlight the stable island chains considered in this Letter. The dots $(G_1)$, triangles $(G_2)$, and stars $(G_3)$ mark the stable orbits corresponding to ($\#$ turns around the center, $\#$ reflections)$=(1,2), (3,8)$, and a David star, shown at the top (cf. Sec.~\ref{s0} in the appendix). The dashed line at $p=p_c$ indicates the critical value of total internal reflection.}
        \label{Fig1}
        \end{figure}

The observed $nkR$-dependent behavior of the decay rates can be explained by a semiclassical model incorporating three effects: \emph{localization}, \emph{Fresnel reflection} (FR), and the \emph{Goos-H\"anchen shift} (GHS). While the first effect occurs in both Hermitian and non-Hermitian systems, the latter two are unique features observed, e.g., in dielectric microcavities or QBs with soft walls. Namely, unlike a single ray, a wave beam impinging on the boundary comprises distributed incident angles, which means that the FR effectively results from an average of these angles. Additionally, because waves are partially transmitted through a dielectric boundary, the incident and reflected positions are shifted with respect to one another. This GHS effect arises from the interference between the incident wave and the phase-shifted reflected wave.
        \begin{figure*}[t]
        \centering
        \includegraphics[width=0.8\textwidth]{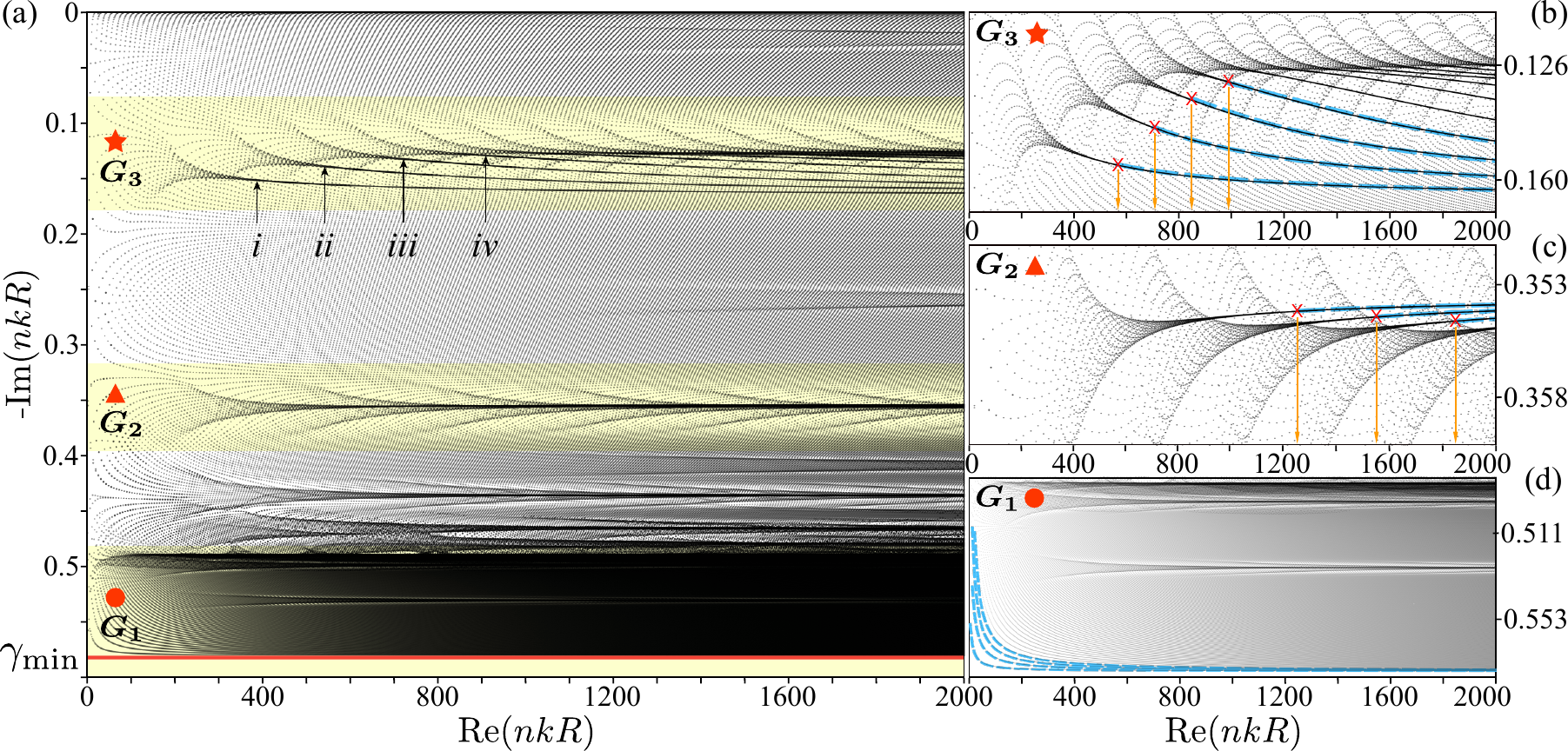}
        \vspace{-0.4cm}
        \caption{(a) Imaginary versus real part of the eigenmodes of the AOM with the boundary shape~\eq{bd} with $\varepsilon=0.05$ and $n=2$. The yellow-shaded regions, labeled by $G_1$, $G_2$, and $G_3$, respectively, are associated with the periodic orbits and stable islands with the same labels and symbols in~\fig{Fig1}. The branches $\{i,ii,iii,iv\}$ are used to illustrate the link between them and POs in Figs.~\ref{Fig3},~\ref{Fig4}, and~\ref{Fig5}. The straight red line indicates the lower bound $\gamma_\mathrm{min}$ of Im$(nkR)$. (b), (c), and (d) show zooms into the branches marked by $G_3$, $G_2$, and $G_1$, respectively, in (a). The thick-dashed blue curves are obtained from~\eq{semik}. The red crosses mark the values of Re$(nkR)$ 570, 710, 850, and 990 in (b) and 1250, 1550, and 1850 in (c) (cf. Sec.~\ref{s6} of the appendix and \fig{Fig6}).}
        \label{Fig2}
        \end{figure*}

\section{The eigenmode spectrum of the AOM and the corresponding classical PSOS\label{Eig}} 
 The AOM under consideration is a quadrupole-shaped two-dimensional dielectric microdisk given in polar coordinates by 
        \begin{align}
        \label{bd}
        r(\theta)/R=r_0\left[1+\varepsilon\cos(2\theta)\right],
        \end{align}
where $\varepsilon$ denotes the deformation parameter, and $r_0=1/\sqrt{1+\varepsilon^2/2}$ ensures that the area $\pi R^2$ of the original circle is preserved. The PSOS of the corresponding quadrupole billiard is defined in terms of Birkhoff coordinates~\cite{birkhoff1927periodic} $(q,p)=(s, \sin\chi)\in[0,L]\times[-1,1]$. Here, the arc length $s$ is measured from the rightmost point of the boundary, $L$ is the perimeter, and $\sin\chi$ is the tangential momentum with $\chi$ denoting the angle between the incident ray and inward-pointing normal to the boundary at $s$ [cf.~\fig{Fig1}]. For a deformation $\varepsilon=0.05$, the PSOS, exhibited in~\fig{Fig1} (cf. Sec.~\ref{s0} in the appendix), consists of a complex mixture of invariant curves, stable resonance chains, and chaotic layers, which renders the analysis of semiclassical phenomena a challenging task.
         
Assuming a harmonic time variation, $\Psi(\mathbf{r},t)\propto \psi(\mathbf{r})e^{-i\omega t}$, the scalar Helmholtz equation of the AOM, i.e. Schr\"odinger equation of the corresponding QB~\footnote{Due to the equivalence of the wave equations we refer to the AOM and corresponding QB on the same level.} is given by 
        \begin{align}
        \label{helm}
        -\nabla^2\psi(\mathbf{r})=n^2k^2\psi(\mathbf{r}),
        \end{align}
subject to dielectric boundary conditions. The refractive index of the AOM is chosen as $n=2$, and for the surrounding air, it equals unity, so the critical angle of total internal reflection is according to Snell's law $p_c=1/n=0.5$ [cf.~\fig{Fig1}]. We focus on electromagnetic wave modes with transverse-magnetic polarization. Then, $\psi$ corresponds to the electric-field component perpendicular to the billiard plane. The modes are obtained by imposing the boundary conditions that $\psi$ and its normal derivative $\partial_\nu\psi$ are continuous across the boundary~\cite{EM_jackson} and that $\psi$ is a purely outgoing wave at infinity. The eigenstates $\{\psi_j,k_j\}$ are complex-valued, $\psi_j,k_j\in \mathbb{C}$, and the eigenmodes $\psi_j$ decay with the rate Im$k$. Since the wave equations are scale-invariant, we use dimensionless wavenumbers $kR$ throughout the analysis.

        \begin{figure}[t]
        \centering
        \includegraphics[width=0.8\columnwidth]{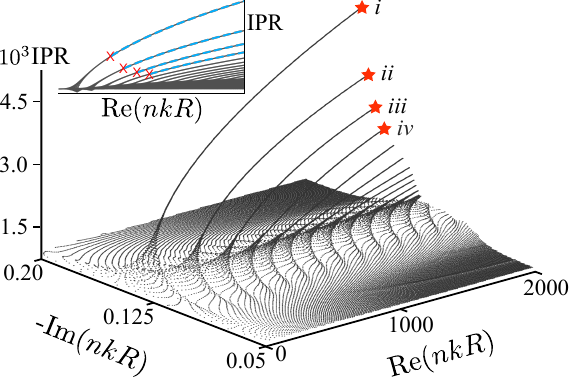}
        \vspace{-0.4cm}
        \caption{Inverse participation ratios of the boundary waves for modes on the   branches $\{i,ii,iii,iv\}$ belonging to $G_3$ in~\fig{Fig2}. The stars mark the values for which wave functions and Husimi functions are shown in~\fig{Fig4}. The dashed blue curves in the inset are obtained from fits of $\mathrm{IPR}^\mathrm{fit}=c_1\sqrt{\mathrm{Re}(nkR)}+c_2$ to the curves yielding, e.g., for the branch $i$ $(c_1,c_2)=(1.13\times 10^{-4},-4.80\times 10^{-4})$. The crosses mark the same values of Re$(nkR)$ as in~\fig{Fig2} (b).}
        \label{Fig3}
        \end{figure}
For the numerical solution of the boundary-value problem, we implemented a boundary-element method (BEM)~\cite{Jan_Wiersig_2003} taking into account the four mirror-reflection symmetry classes~\cite{Spiridonov:17} and incorporating M\"uller's BEM~\cite{Muller0}, Kress's quadrature scheme~\cite{KRESS1991229}, the block Sakurai-Sugiura method~\cite{IKEGAMI20101927}, and Veble's perturbative approach~\cite{Veble_2007} as complementary methods to enhance efficiency and accuracy. In~\fig{Fig2}, we plot Im$(nkR)$ versus Re$(nkR)$. We refer the readers to Sec.~\ref{s0} in the appendix for details. Within the range Re$(nkR)\in[0,2000]$, we obtained $\sim250,400$ eigenstates for each symmetry class. A comparison with the expected number of eigenmodes, as given by Weyl's law~\cite{PhysRevE.78.056202,PhysRevE.83.036208}, indicates that, at most, 20 extremely high-quality modes with $|$Im$(nkR)|\simeq 0$ are missing.
    
Figure~\ref{Fig2} reveals various intriguing structures as Re$(nkR)$ increases. First, the well-known lower bound, $\mathrm{Im}(nkR) > \frac{R}{\xi_\mathrm{min}} \mathrm{ln} \frac{n-1}{n+1} = \gamma_\mathrm{min}$ ~\cite{PhysRevE.78.056202,PhysRevE.83.036208}, marked by a red line, is validated. Here, $\xi_\mathrm{min}$ denotes the length of the shortest periodic orbit (PO) with $\chi=0$, corresponding to the vertical diameter orbit denoted by $G_1$ in \fig{Fig1}, where $\xi_\mathrm{min} = \frac{2R(1-\varepsilon)}{\sqrt{1+\varepsilon^2/2}}$, yielding $\gamma_\mathrm{min} \approx -0.5786$. The most remarkable finding is the occurrence of multiple sequences of branches of Im$(nkR)$ merging into curves that change little with increasing Re$(nkR)$. We show in the following that each branch is associated with a PO, such as $G_1$, $G_2$, and $G_3$ in \fig{Fig2}.
    
\section{Semiclassical approach for the determination of the decay rates\label{Semicl}}    
In the following, we present a detailed analysis of the decay rates using the branches associated with the PO $G_3$ as an example. To identify the corresponding modes, we first computed the inverse participation ratio (IPR) of the boundary wave functions $\psi^b(s)$, defined as IPR$ = \frac{\sum_j |\psi^b(s_j)|^4}{\left(\sum_j |\psi^b(s_j)|^2\right)^2}$, which provides a measure of localization~\cite{PhysRevB.83.184206}. Note that, we discretized the arc-length parameter $s$, choosing the distance between the points $s_j$ to be smaller than one quarter of the wavelength.
        \begin{figure}[t]
        \centering
        \includegraphics[width=0.8\columnwidth]{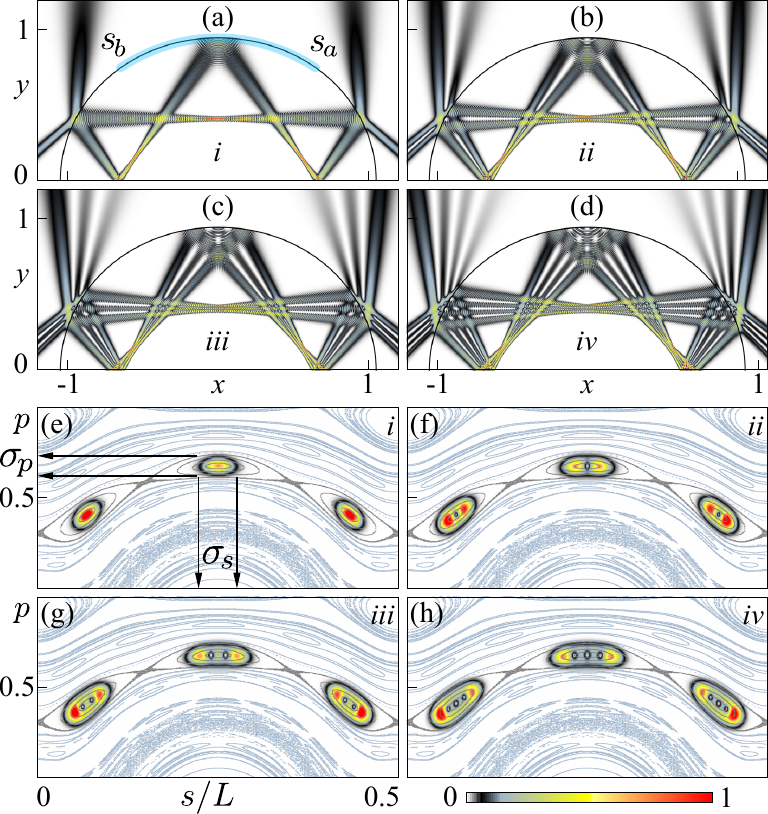}
        \vspace{-0.4cm}
		\caption{Wave-function intensity distributions of the modes for the branches $i$ (a), $ii$ (b), $iii$ (c), and $iv$ (d) in~\fig{Fig3}. In (a) the range of $s\in [s_a,s_b]$ taken into account in~\fig{Fig5} is marked by a blue curve. In (e), (f), (g), and (h) the associated Husimi distributions are shown. These are overlayed with the PSOS to illustrate the connection of the branches to islands, that is, POs.  The color code is provided at the bottom. The quantities $\sigma_s$ and $\sigma_p$ given in (e) illustrate the localization width of the modes in (a) along the $s$ and $p$ directions.}
        \label{Fig4}
        \end{figure}
We demonstrate in~\fig{Fig3} that the IPRs of consecutive converging branches of Im$(nkR)$ grow according to the scaling relation IPR $\propto \sqrt{\mathrm{Re}(nkR)}$ [cf. the inset of~\fig{Fig3}]. We picked for each of these branches one mode with Re$(nkR) \approx 2000$ (marked by red stars in~\fig{Fig3}) and show in~\fig{Fig4} the intensity distributions of their wave functions, i.e., electric field strength and the corresponding (inside-incoming) Husimi distributions~\cite{Husimi1940,M.Hentschel_2003}, which are also referred to as quantum PSOS~\cite{Baecker2004}. The intensities reveal a strong localization along wave beams of finite-width that are on or close to the PO $i$ and its repetitions $l=1,2,3$ for $ii,\ iii,\ iv$, respectively. Here, the mirror symmetry of the AOM about the $x$-axis is considered. Furthermore, for visualization purpose, we display the symmetrized Husimi distributions, $\mathcal{H}(s,p) = [\mathcal{H}(s,p) + \mathcal{H}(s,-p)]/2,\ p \ge 0$.

       \begin{figure}[t]
       \centering
       \includegraphics[width=0.8\columnwidth]{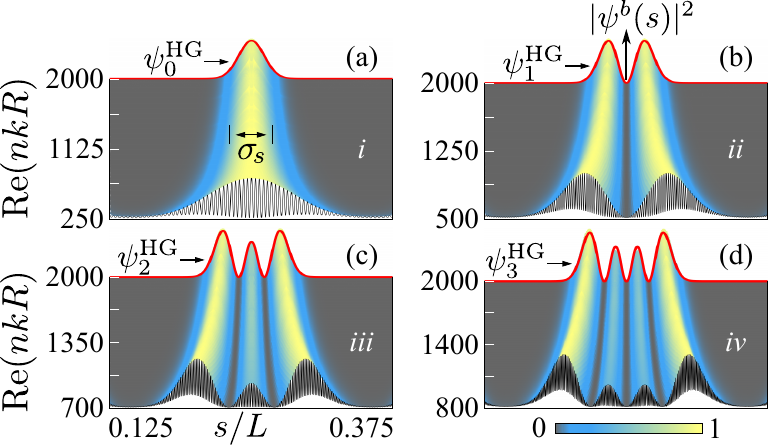}
       \vspace{-0.4cm}
		\caption{Envelopes of the intensities of the boundary waves, $|\psi^b(s)|^2$, as function of Re$(nkR)$, for branches $i$ (a), $ii$ (b), $iii$ (c), and $iv$ (d) in~\fig{Fig2}. The plot range, $s\in [0.125L,0.375L]$ is depicted as blue curve in~\fig{Fig4}(a). The red curves show $\psi_l^\mathrm{HG}(s)$ at Re$(nkR)=2000$ for $l=0,1,2,3$, respectively. In (a), $\sigma_s$ indicates the Gaussian width of $\psi_l^\mathrm{HG}(s)$.}
\label{Fig5}
\end{figure}
To understand the mechanism underlying the Re$(nkR)$-dependence of the IPRs, we analyze the intensity distributions of the boundary wave functions, $|\psi^b(s)|^2$ in the range $s/L\in[s_a,s_b]=[0.125,0.375]$ marked by a blue curve in~\fig{Fig4}(a), exhibited in~\fig{Fig5} for the four branches $i,\ ii,\ iii,$ and $iv$. Their envelopes agree well with the $l$th-order Hermite-Gauss function (red curves) centered at $s_c/L=0.25$, $\psi^\mathrm{HG}_l(s)=\sqrt{\frac{\mathcal{N}}{2^ll!\sqrt{\pi}}} \mathrm{H}_l\left[\frac{s-s_c}{\sigma_s}\right]e^{-\frac{1}{2\sigma_s^2}(s-s_c)^2}$, for $l=0,1,2,3$, respectively, as illustrated for Re$(nkR)=2000$ in~\fig{Fig5}. Here the factor $\mathcal{N}$ scales Max$[|\psi_l^\mathrm{HG}(s)|^2]$ to unity, H$_l$ denotes the Hermite polynomial and $\sigma_s$ the Gaussian width. We find that $\sigma_s\propto 1/\sqrt{\text{Re}(nkR)}$, which is in accordance with the Re$(nkR)$-dependence of the IPRs observed in~\fig{Fig3}. Furthermore, as these modes are localized on stable islands in the PSOS, the width in the momentum direction $\sigma_p$ [cf.~\fig{Fig4}(e)] exhibits the same behavior, $\sigma_p\propto 1/\sqrt{\text{Re}(nkR)}$. Thus the area covered by these modes in the PSOS contracts as $\sigma_s\sigma_p\propto1/\mathrm{Re}(nkR)$, confirming the semiclassical prediction, $h_\mathrm{eff}\simeq 1/\mathrm{Re}(nkR)$~\cite{PhysRevE.63.056203}, for the effective Planck constant, which provides a measure for the resolution in phase space. The enhancement of localization with increasing Re$(nkR)$ explains the formation of branches. Namely, the associated island with constant phase space area $\oint p \, \deri q = \Omega$ can accommodate an increasing number of states ($\sim \Omega/\h$) corresponding to a sequentially higher order $l$, with $l\to\infty$ as the classical limit, $\h\to 0$, is approached. This process is consistent with the existence of an infinite number of concentric tori within the island in classical phase space.
    
Next, we outline the derivation of a semiclassical model for the decay rates in the SR. The FR coefficient is given by~\cite{PhysRevE.78.056202,Luhn_2020}.
        \begin{align}
        \label{semir}
		\mathcal{F}(m,n,x)=&\frac{\frac{H_{m-1}^{(1)}(x)}{H_m^{(1)}(x)}-n\frac{H_{m-1}^{(1)}(n x)}{H_m^{(1)}(n x)}}{\frac{H_{m-1}^{(1)}(x)}{H_m^{(1)}(x)}-n\frac{H_{m-1}^{(2)}(n x)}{H_m^{(2)}(n x)}},
        \end{align}
with $x=\mathrm{Re}(k\beta)$, $\beta$ denoting the radius of curvature at the reflection point and $H_m^{(\zeta)}$ the $m$th-order Hankel function of the first ($\zeta$=1) and second ($\zeta$=2) kind. An estimate for the order $m$ can be obtained from the EBK quantization, $m\approx\mathrm{Re}(nk\beta) p$~\cite{Keller1960,JUNockel}. In Sec.~\ref{s8}  in the appendix more details are provided and in Fig.~\ref{fig:semiR} (a) $\mathcal{Z}(m,n,x)=\vert\mathcal{F}(m,n,x)\vert^2$ is shown for the case $(n,\beta)=(2,R)$ as a function of Re$(nkR)$.

A semiclassical approximation of the decay rate Im$(nkR)$ is obtained from the relation~\cite{PhysRevE.78.056202,PhysRevE.65.045603}
        \begin{align}
        \label{semik}
        \mathrm{Im}(nkR)=\frac{\sum_{\tau=1}^N\ln\left\{\mathcal{Z}\left[\mathrm{Re}(nk\beta_\tau) p_\tau,n,\mathrm{Re}(k\beta_\tau)\right]\right\}}{2/R\sum_{\tau=1}^N\xi_\tau},
        \end{align}
where the sum is over the contributions from each of $N$ reflection points $(s_{\tau},p_{\tau})$ of a single trajectory consisting of $N$ segments of length  $\xi_\tau =\Vert\mathbf{r}(s_\tau)-\mathbf{r}(s_{\tau-1})\Vert$ in configuration space [cf.~\fig{Fig1}]. To generate the ray dynamics taking into account the Re$(nkR)$ dependence of the localization and the shift of the reflection points with each pass along a PO, like that observed in~\fig{Fig4} with increasing $0\leq l\leq 3$, we chose $l$ and Re$(nkR)$-dependent initial values $(s_0,p_0)_l=\left(S_0,P_0+\sqrt{\frac{\kappa_0+l \kappa}{\mathrm{Re}(nkR)}}\Delta P\right)$. Here $(S_0,P_0)$ denotes the initial conditions of the associated classical PO, i.e., $(S_0,P_0)=(0.25L,0.57)$ for the branches belonging to $G_3$ [cf.~\fig{Fig4}]. The other parameters were chosen as $\left\{\Delta P,\kappa_0,\kappa\right\}=\left\{0.025,10,174\right\}$.

Furthermore, we introduce another essential effect, the GHS, that needs to be incorporated into~\eq{semik}, as demonstrated in Sec.~\ref{s7} in the appendix. This shift induces a displacement $\delta s$ of POs and generates an increment $\delta p$ to $p$ at each reflection point, resulting in the transformation $(s_{\tau},p_{\tau})\mapsto (s_{\tau}+\delta s,p_{\tau}+\delta p)$. Here, $\delta p$ is a function of the induced shift $\delta s$ at $p_\tau$~\cite{PhysRevE.82.026202}, $\delta p[\delta s(p_\tau)]=\frac{\sqrt{1-p_\tau^2}}{\beta}\delta s(p_\tau)$. In the calculation of Im$(nkR)$ using~\eq{semik}, we neglect the shift $\delta s$ in $s_\tau$ and assume that only $p_\tau$ changes, $p_\tau\to p_\tau+\delta p$. Previous works~\cite{PhysRevLett.120.093902,PhysRevE.100.042219,PhysRevA.96.023848,PhysRevE.94.022202} have verified that this assumption holds well. Furthermore, we can also neglect the contribution of $\delta s$ to $\xi_\tau$ because $\xi_\tau\gg\delta s$. To compute the GHS $\delta s(p)$, we use the formula derived in~\cite{PhysRevA.93.023801}, which has no divergences at the critical angle,
        \begin{align}
        \label{goos}
        \delta s(p)=R\sqrt{\frac{w \pi p /\left(1-p^2\right)^\frac{3}{2}}{2^\frac{3}{2} \mathrm{Re}(nkR)}}\mathcal{I}\left[\frac{1-n^2p^2}{2^\frac{5}{2}n^2p\sqrt{1-p^2}}w\mathrm{Re}(nkR)\right],
        \end{align}
with $\mathcal{I}(z)=e^{-z^2}\sqrt{|z|}\left[I_{-\frac{1}{4}}\left(z^2\right)-\frac{z}{|z|}I_\frac{1}{4}\left(z^2\right)\right]$. Here, we chose for the Gaussian width $w=\eta/\sqrt{\mathrm{Re}(nkR)}$, and $I_a$ refers to the first kind of modified Bessel function of order $a$. We set $\eta=1/\sqrt{2}$ for the branches $\{i,ii,iii,iv\}$. In Fig.~\ref{fig:semiR} (b) in the appendix $k\delta s(p)$ is plotted versus $p$ and Re$(nkR)$. The decay rates obtained by \eq{semik} for $G_3$ are shown as thick-dashed blue curves in \fig{Fig2}(b). The agreement with the BEM results is very good for values of Re$(nkR)$ beyond the points marked by red crosses. To corroborate the validity of our model, we extended the analysis to the branches marked by $G_1$ and $G_2$ in~\fig{Fig2}, finding that the corresponding modes are localized on the POs $G_1$ and $G_2$ in \fig{Fig1} (cf. Sec.~\ref{s0} and also S1 in~\cite{suppl} for additional examples). Equation~(\ref{semik}) provides the thick-dashed blue curves in Figs.~\ref{Fig2}(c) and (d), which again lie on top of the branches formed by the corresponding decay rates.
        \begin{figure}[t]
        \centering
        \includegraphics[width=0.9\columnwidth]{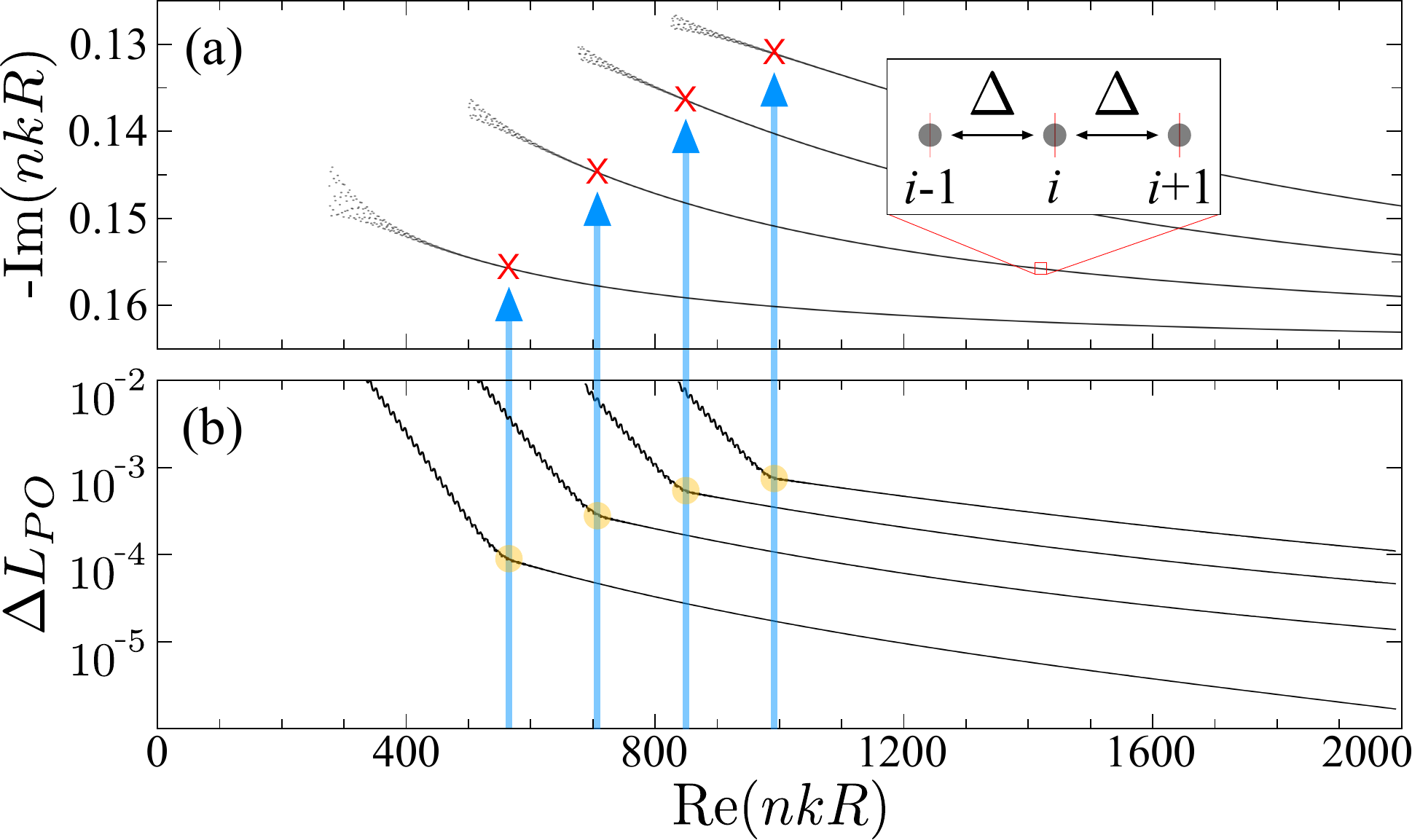}
        \vspace{-0.4cm}
		\caption{The decay rates (a) and deviations $\Delta L_{PO}$ (b) for the modes along the branches in~\fig{Fig2}(b), associated with PO $G_3$ of length $L_{\text{exact}} = 10.42647950$. In (b), kinks are observed pointing at significant slope changes. Those of the neighboring branches, corresponding to $l=0,1,2,3$ transverse repetitions of the PO $G_3$, are equidistant in Re$(nkR)$, as expected~\cite{Tureci:02}.}
        \label{Fig6}
        \end{figure}

The values of Re$(nkR)$, marked by crosses in Figs.~\ref{Fig2}(b) and (c), and in the inset of \fig{Fig3}, were determined based on the observation that, even in the deep SR, the wave functions—such as those in~\fig{Fig4}—are not perfectly localized on the associated PO but instead on finite-width wave beams close to it. To quantify this, we calculated the deviation between the exact length, $L_{exact}$, of the underlying PO and the length of the corresponding wave beam, obtained from the free-spectral-range (FSR)~\cite{PhysRevE.83.036208}. The FSR is given by the difference $\Delta(k_i) = \text{Re}[n(k_i - k_{i-1})R]$, where $k_{i-1}$ and $k_i$ represent the wavenumbers of consecutive modes along a branch [see inset of Fig.~\ref{Fig6}(a)]. This yields the effective PO length of the wave beam at wavenumber $k_i$ as $L_{\text{FSR}} = \frac{2\pi}{\Delta(k_i)}$ (cf. Sec.~\ref{s6} in the appendix). Figure~\ref{Fig6}(b) shows the deviation $\Delta L_{PO} = \vert L_{exact} - L_{\text{FSR}} \vert$ as a function of Re$(nkR)$. The curve reveals a kink, indicating a significant change in slope. We conclude that for Re$(nkR)$ values below the kink, the effective Planck constant, $h_{eff} \approx 1/\text{Re}(nkR)$, is too large to resolve the fine structure of the island chain. However, for values beyond the kink, the modes can be considered localized along the PO, consistent with the assumptions in our semiclassical model. In this Re$(nkR)$ range, observed deviations are primarily attributable to tunneling processes~\cite{PhysRevLett.104.114101}.

\section{Conclusions\label{Concl}}   
The numerical analysis of the Re$(nkR)$ dependence of the decay rates up to Re$(nkR)\simeq 2000$ reveals a series of converging branches, that have not been observed, the reason being that we go far beyond the Re$(nkR)$ range that has been explored before. One exception is Ref.~\cite{PhysRevLett.129.193901}, where the focus was on the fluctuation properties of the modes. The observed intriguing branching arises from the Re$(nkR)$-dependent localization of higher-order modes in stable resonance chains in mixed phase space. We demonstrate that the decay rates can be employed to resolve the island structure. Furthermore, we propose an effective model for the decay rates of those modes, which incorporates localization, semiclassical Fresnel reflection, and the Goos-H\"anchen shift, and provide a lower bound of Re$(nkR)$, beyond which it is applicable and the correspondence between the decay rates and the associated classical periodic orbit holds.

Our discovery enhances our understanding of quantum chaos, as it provides another signature of the classical mixed regular-chaotic dynamics in terms of decay rates of optical modes in the partially opened quantum system. In addition, it sheds light on various optical and photonics phenomena in the SR. This includes mode formation in optical microresonators and semiconductor lasers, as well as light propagation dynamics in optical waveguides~\cite{vahala,RevModPhys.87.61,postech}. Furthermore, our study can be extended to mesoscopic physics, offering valuable insights into transport phenomena in quantum-dot-based devices. Finally, this work provides a reliable foundation for future studies in quantum chaos in non-Hermitian systems, including the connection between complex-valued energy level statistics and the degree of chaoticity, which we are currently investigating for various deformations $\varepsilon$. 

The financial support from Institute for Basic Science (IBS-R024-D1) is acknowledged.\\

\bibliographystyle{apsrev4-2}    

%**********************************
%\bibliography{reference}
%**********************************


\begin{appendix}
\begin{widetext}
\renewcommand{\theequation}{A\arabic{equation}}
\renewcommand{\thefigure}{A\arabic{figure}}
\setcounter{figure}{0}
\section{Wave intensities, Husimi distribution, and IPRs for the groups $G_1$ and $G_2$ in Fig.~\ref{Fig1}\label{s0}}
In Fig.~\ref{fig:0} we show examples for the wave intensity $|\psi(x,y)|^2$, the Husimi distribution $\mathcal{H}(s,p)$, and the IPRs for the groups $G_1$ and $G_2$.
\begin{figure}[!h]
   \centering
    \includegraphics[width=0.5\textwidth]{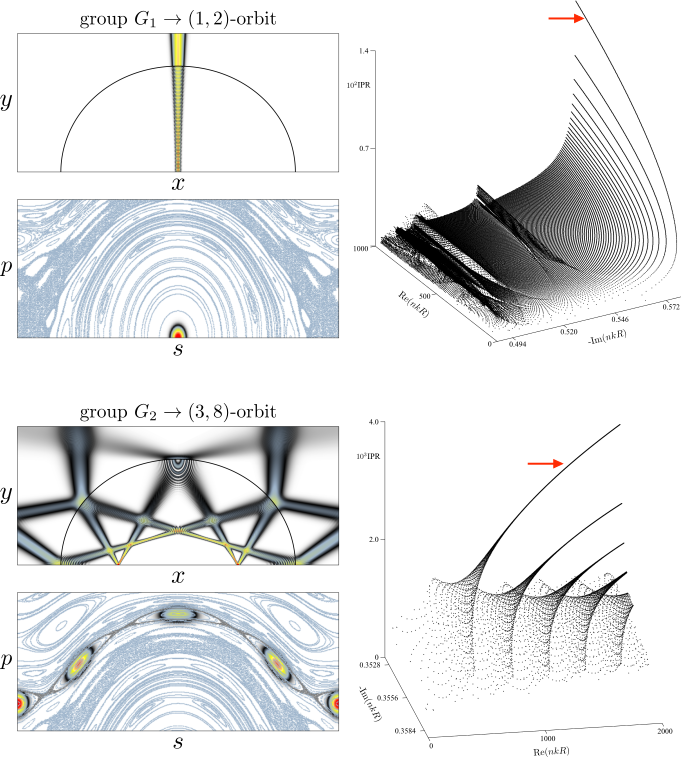}
    \caption{The wave intensity $|\psi(x,y)|^2$, the Husimi distribution $\mathcal{H}(s,p)$, and the inverse participation ratio (IPR) for the groups, $G_1$ (upper three figures) and $G_2$ (lower three figures) discussed in the main text. Their $k$ values are on the branches exhibited by the decay rates and indicated by arrows in the corresponding IPR plot.}
    \label{fig:0}
\end{figure}
Figure~\ref{fig:1} illustrates in more detail than in Figs.~\ref{Fig1} and~\ref{Fig2} the correspondence between Im$(nkR)$ versus Re$(nkR)$ and the structurei of the PSOS.  The groups $\{G_1,G_2,G_3\}$ are the ones discussed in the main text and Fig.~\ref{fig:0}. In Fig.~\ref{fig:2}, this correspondence is explicitly demonstrated by examining waves and Husimi distributions of modes belonging to those groups. The $nkR$ values of the high-Q mode belonging to \quoting{1} and \quoting{2} are shown in Figs.~\ref{fig:3} and~\ref{fig:4}. Owing to the extremely small decay rate, $|\text{Im}(nkR)|<10^{-12}$, the modes for the islands above the ones of the groups \quoting{1} are not precisely identifiable. 
\begin{figure}[!h]
   \centering
    \includegraphics[width=0.7\textwidth]{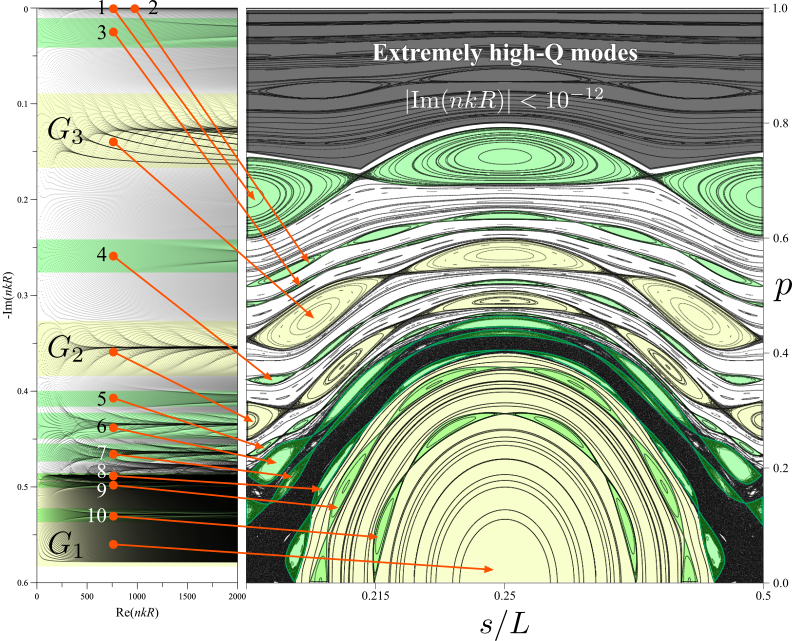}    
	\caption{Left: $\text{Im}(nkR)$ vs. $\text{Re}(nkR)$ of modes. Right: Phase space of ray dynamics. The converging families of $\text{Im}(nkR)$ marked as $\{G_1,G_2,G_3,1,2,3,4,5,6,7,8,9,10\}$ correspond to the islands in phase space indicated by the arrows. The groups $\{G_1,G_2,G_3\}$ are the ones discussed in the main text and shown Fig.~\ref{fig:0}.} 
\label{fig:1}
\end{figure}
\begin{figure}[!h]
   \centering
    \includegraphics[width=0.7\textwidth]{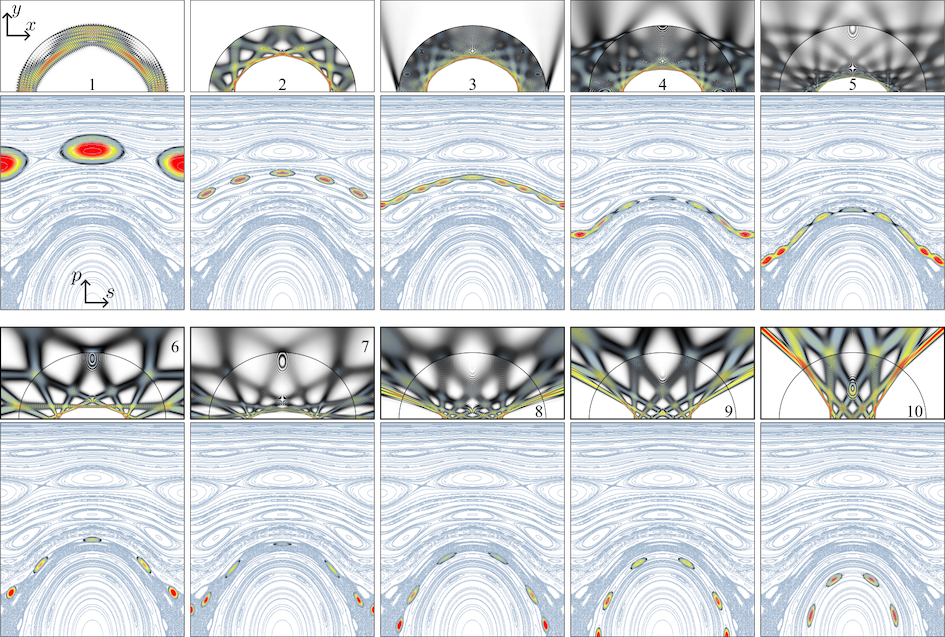}
    \caption{The first [second] and third [fourth] rows represent the wave intensity $|\psi(x,y)|^2$ [Husimi distribution $\mathcal{H}(s,p)$] of the selected modes belonging to the groups $\{1,2,3,4,5,6,7,8,9,10\}$ in Fig.~\ref{fig:1}. The $nkR$ values of the selected modes corresponding to \quoting{1} and \quoting{2} are indicated by arrows in Figs.~\ref{fig:3} and~\ref{fig:4}, respectively.}
    \label{fig:2}
\end{figure}
\begin{figure}[!h]
   \centering
    \includegraphics[width=0.45\textwidth]{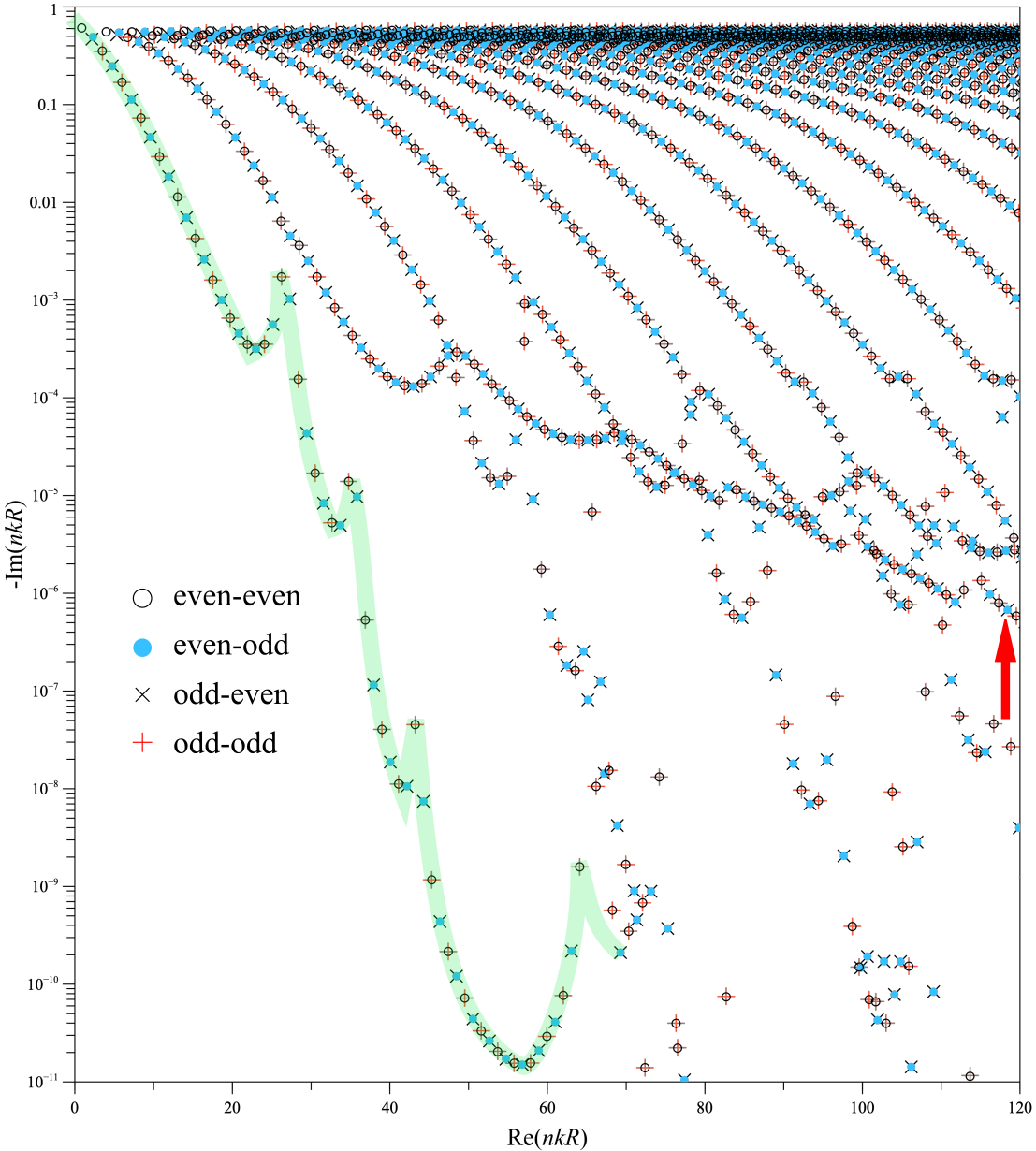}
    \caption{Logarithmic plot of $\text{Im}(nkR)$ versus $\text{Re}(nkR)$ for the group \quoting{1} in Fig.~\ref{fig:1}. The symbols of open circle, filled circle, cross, and plus respectively represent the mirror-reflection symmetry class of $(M_x,M_y) = (\text{even},\text{even})$, $(\text{even},\text{odd})$, $(\text{odd},\text{even})$, and $(\text{odd},\text{odd})$: $M_x$ $(M_y)$ denotes the mirror-reflection symmetry operation with respect to the $x(y)$-axis, and \quoting{even} and \quoting{odd} stand for \quoting{$+1$} and \quoting{$-1$} in which $M_x \psi(x,y) = \pm \psi(x,-y)$ and $M_y \psi(x,y) = \pm \psi(-x,y)$. The mode corresponding to the wave and the Husimi distribution of the group \quoting{1} shown in Fig.~\ref{fig:2} is indicated by the arrow. The highlighted points by the thick curve manifest the spikes of the Q-spoiling phenomenon explained by the resonance-assisted dynamical tunneling effects [see, e.g., Kullig et al., Phys. Rev. E \textbf{94}, 022202 (2016)].}
    \label{fig:3}
\end{figure}
\begin{figure}[!h]
   \centering
\includegraphics[width=0.45\textwidth]{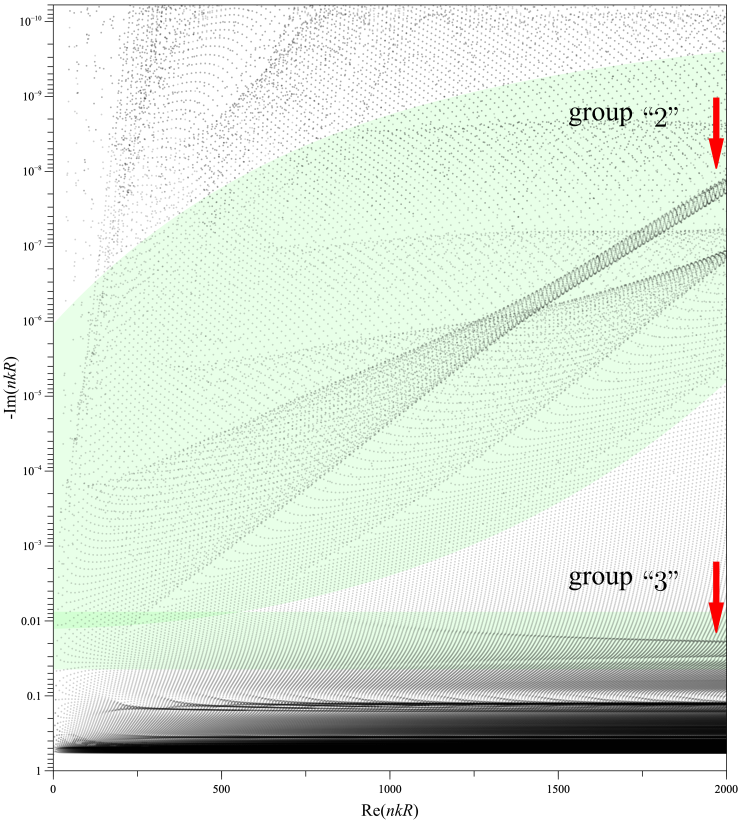}
\includegraphics[width=0.45\textwidth]{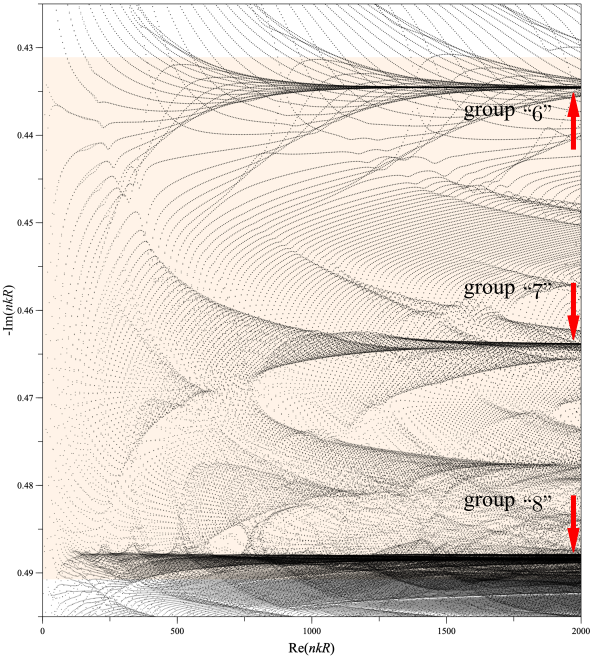}	
    \caption{Left: Logarithmic plot of $\text{Im}(nkR)$ versus $\text{Re}(nkR)$ for the groups \quoting{2} and \quoting{3} in Fig.~\ref{fig:1}. The modes corresponding to the wave and the Husimi distribution of the groups \quoting{2} and \quoting{3} shown in Fig.~\ref{fig:2} are indicated by the arrows. Right: Modes corresponding to the largest chaotic layer enclosing the largest island of $G_1$. In Fig.~\ref{fig:1}, the darkest region immediately outside the largest island of $G_1$ in phase space is the largest chaotic layer. This chaotic layer embeds recognizable three island structures corresponding to the groups \quoting{6}, \quoting{7}, and \quoting{8}. As is expected, the structures of $nkR$ in the shaded region between the group \quoting{6} and \quoting{8} are notably complicated. The mode corresponding to the wave and the Husimi distribution of the groups \quoting{6}, \quoting{7}, and \quoting{8} shown in Fig.~\ref{fig:2} is indicated by the arrow.}
\label{fig:4}
\end{figure}
The right part of Fig.~\ref{fig:4} shows modes in the largest chaotic layer enclosing the largest island of $G_1$.
\begin{figure}[!h]
   \centering
    \includegraphics[width=0.5\textwidth]{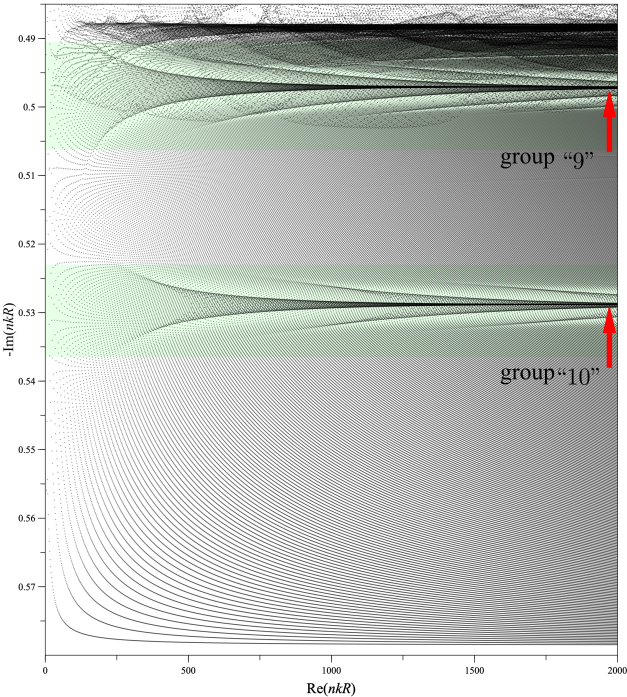}
    \caption{Zoomed-in view of the modes associated with group $G_1$ shown in Fig.~\ref{fig:1}. Sub-group structures \quoting{9} and \quoting{10} correspond to satellite islands within the largest island of $G_1$ [as indicated by arrows in the phase space plot shown in Fig.~\ref{fig:1}]. The modes for groups \quoting{9} and \quoting{10}, analyzed in Fig.~\ref{fig:2}, are indicated by arrows.}
    \label{fig:6}
\end{figure}
\begin{figure}[t!]
   \centering
    \includegraphics[width=0.7\textwidth]{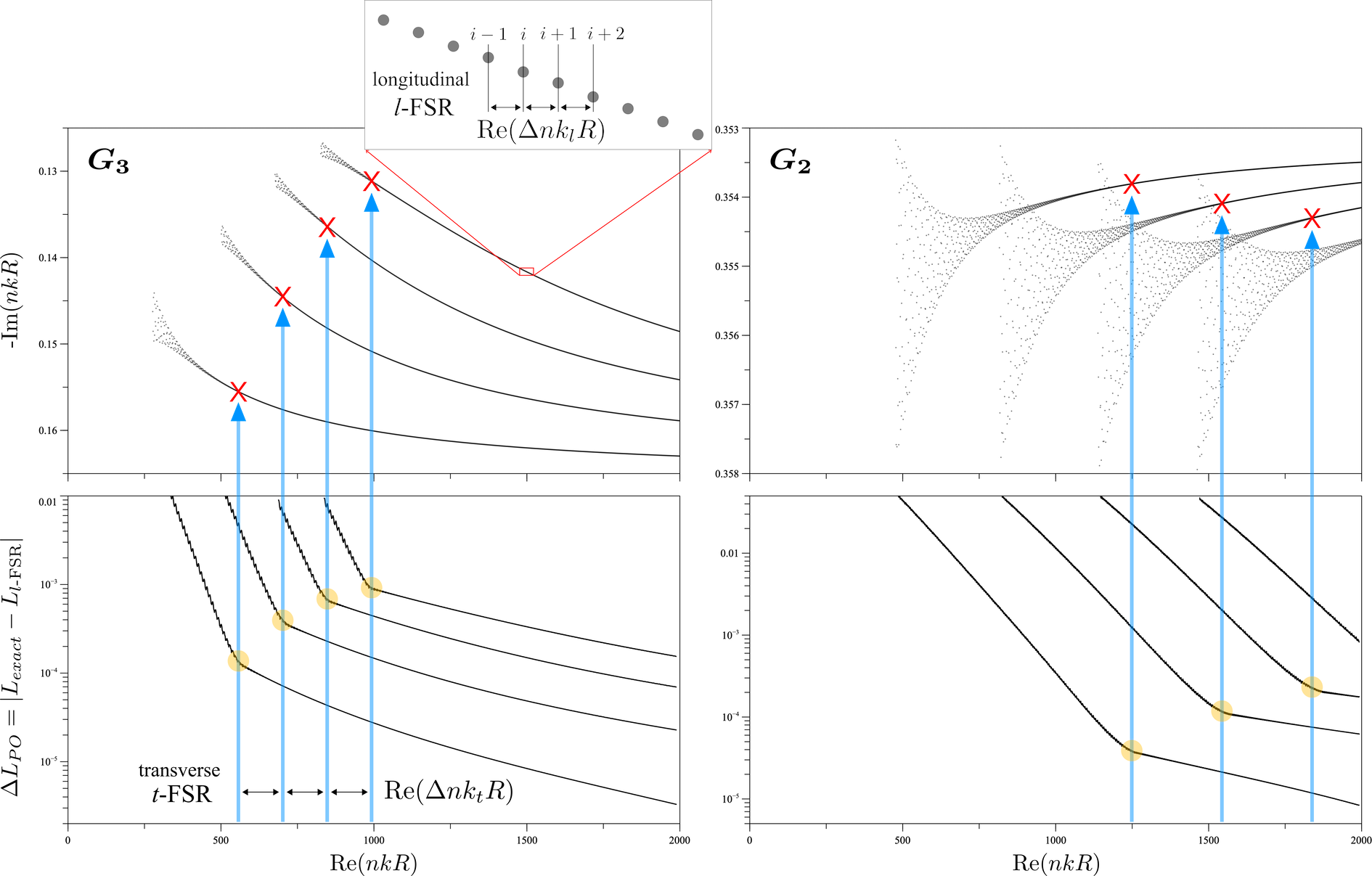}
    \caption{Localization transition of David star [Left: $G_3$] and $(3,8)$-orbit [Right: $G_2$] modes identified by the deviation between the exact periodic orbit length and the one obtained from the longitudinal free-spectral-range ($l$-FSR), Re$(\Delta nk_lR)$. In the bottom figures, we observe that the deviation $\Delta L_{PO} = |L_{exact} - L_{l\text{-FSR}}|$, plotted as a function of Re$(nkR)$, shows periodic slope change points. These points form equidistant sequences which is the transverse free-spectral-range ($t$-FSR), Re$(\Delta nk_tR)$, corresponding to the higher-order modes, discussed, e.g., in Figs. 4 and 5 of the main text for the David star modes.}
    \label{fig:7-1}
\end{figure}

\section{Transition points of localization in the island chain}\label{s6}	
The localization transition of the obtained modes belonging to the individual branches can be characterized by computing the deviation between the exact periodic orbit length and the length derived from the longitudinal free-spectral-range ($l$-FSR). Explicitly, the $l$-FSR is obtained by measuring $\text{Re}(\Delta nk_l R)_i=\text{Re}(nk_iR-nk_{i-1}R)$ for the modes belonging to the same branches [see the inset of the upper-left figure in Fig.~\ref{fig:7-1}], as a function of $\text{Re}(nk_iR)$. From the obtained $l$-FSR, we can deduce the estimated periodic orbit length of the modes, as follows:
\begin{align}
    L_{l\text{-FSR}}=\frac{2\pi}{\text{Re}(\Delta nk_l R)}.
\end{align}
Then, the deduced periodit orbit length is compared with the exact periodic orbit $L_{exact}$ to compute the mutual deviation, as follows:
\begin{align}
  \Delta L_{PO} = |L_{exact} - L_{l\text{-FSR}}|
\end{align}
This deviation reveals a transition point in Re$(nkR)$, marked by a distinct change in slope, as we can see in the bottom figures in Fig.~\ref{fig:7-1}. In the figure, the used values of $L_{\text{exact}}$ are $10.42647950$ for $G_3$ and $14.85876985$ for $G_2$. These values are extracted from the ray dynamics associated with the stable fixed point located at the center of the island chains corresponding to the David star-$(G_3)$ and $(3,8)$-$(G_2)$ orbits.

The interpretation of this slope change is as follows. Prior to the transition point, the modes cannot be considered completely localized in the island chain, as the effective Planck constant, $h_{eff} \simeq 1/\text{Re}(nkR)$, is too large to resolve the fine structure of the island chain. Consequently, the modes are influenced not only by the central island chains but also by other nearby modes associated with the surrounding phase space structure. Namely, many altered periodic orbits are involved. After the transition point, the modes can be considered localized in the central island chain and predominantly governed by it, meaning that only a single periodic orbit corresponds to the localized modes. In this regime, the observed deviation can be attributed primarily to tunneling processes.

Note that the identified transition point of localization reveals another equidistant sequence as well [$\text{Re}(\Delta nk_t R)$; the bottom-left figure in Fig.~\ref{fig:7-1}], which corresponds to the transverse free-spectral-range ($t$-FSR) associated with the higher-order transverse modes illustrated, e.g., in Figs. 4 and 5 of the main text. In addition, the size of the island chain corresponding to the $(1,2)$ orbit is quite large, leading to immediate localization at very small $\text{Re}(nkR)$. In this case, the transition point is unidentifiable for the low-order transverse modes.

\section{Semiclassical Fresnel law and Goos-H\"anchen shift}\label{s8}
\begin{figure}[t!]
   \centering
    \includegraphics[width=0.5\textwidth]{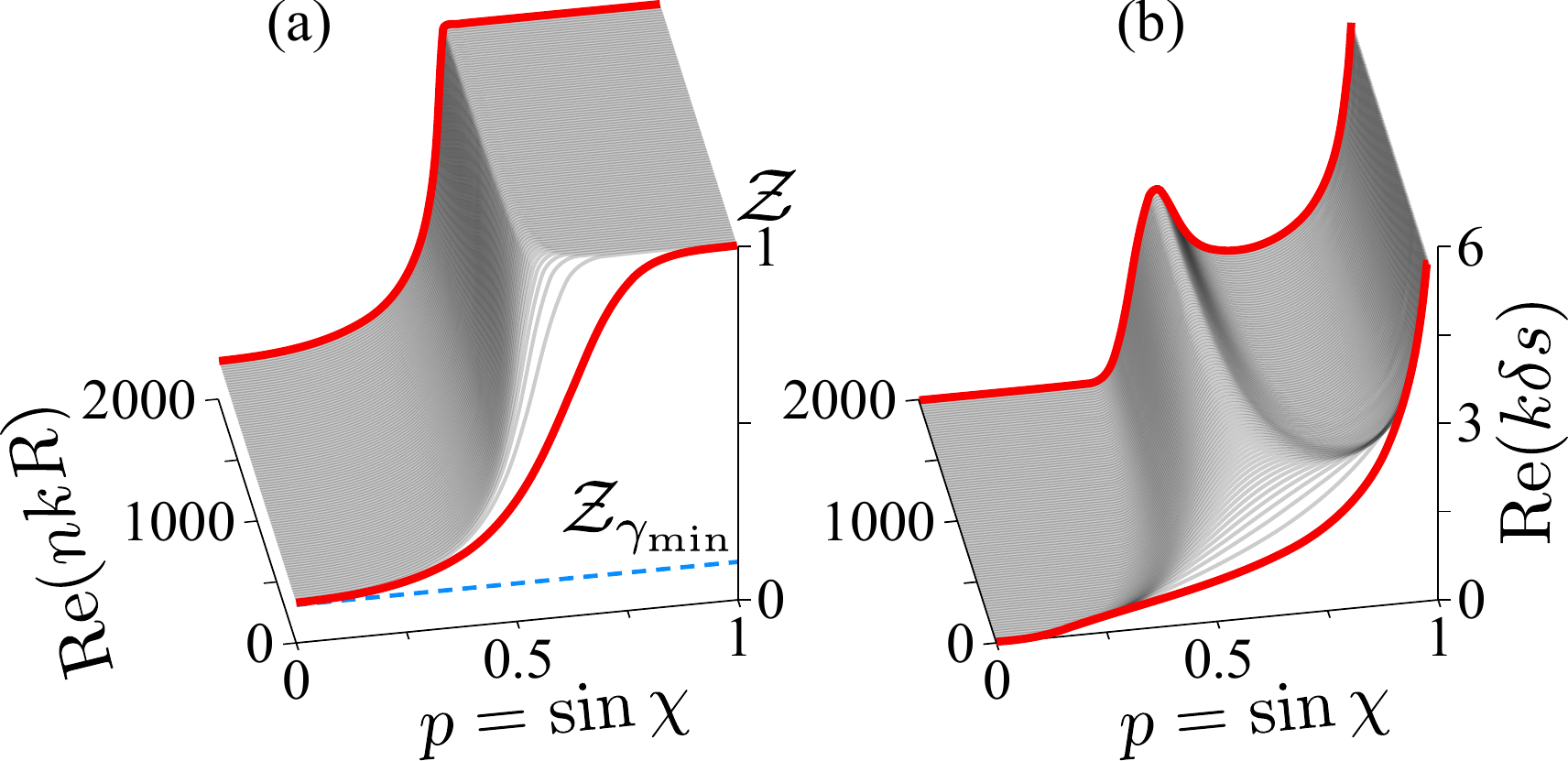}
    \caption{(a) Squared modulus of the semiclassical Fresnel reflection coefficient obtained from Eq. (3) in the main text with $n=2$ and $\beta=R$ as a function of $p$ and Re$(nkR)$. The dashed line at $\mathcal{Z}_{\gamma_\mathrm{min}}$ indicates the minimum of Fresnel reflection at normal incidence. (b) The wavenumber-rescaled Goos-H\"anchen shift obtained with Eq. (5) in the main text, as a function of $p$ and Re$(nkR)$. The two limiting cases of Re$(nkR)\sim0$ and $\sim2000$ are highlighted by thick curves.}
    \label{fig:semiR}
\end{figure}

In~\fig{fig:semiR}(a), we show the squared modulus of the Fresnel coefficient obtained from Eq. (3) for the case of $(n,\beta)=(2,R)$ as a function of $p$ and Re$(nkR)$. It is shown that while $\mathcal{Z}$ smoothly interpolates between the minimum $\mathcal{Z}_{\gamma_\mathrm{min}}=[(1-n)/(1+n)]^2\approx0.11$ and the maximum $\mathcal{Z}=1$ when Re$(nkR)$ is sufficiently small, $\mathcal{Z}$ exhibits a sharp transition point at the total internal reflection point, i.e., $\mathcal{Z}=1$ for $p>p_c=1/n=0.5$ [cf. \eq{se1}] for large  Re$(nkR)$. In~\fig{fig:semiR}(b), we examine the Re$(nkR)$-dependent $\delta s(p)$ rescaled with Re$(k)$, confirming that $\delta s$ stays finite at the critical value $p_c=0.5$ without divergence.

\begin{figure}[t!]
   \centering
    \includegraphics[width=0.5\textwidth]{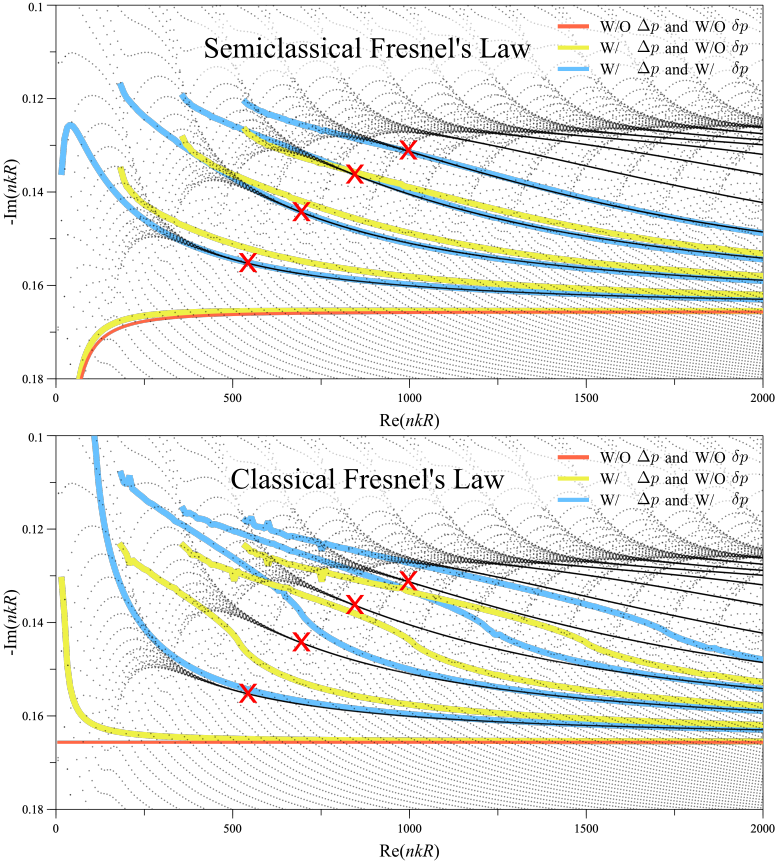}
    \caption{The thick curves show the Re$(nkR)$-dependence of Im$(nkR)$ obtained from Eq. (4) in the main text on top of the black dots resulting from the numerical solutions of Eq. (2) in the main text. The curves in upper panel are based on the semiclassical reflection given by Eq. (4) in the main text, while the ones exhibited in the lower panel are based on the conventional classical reflection given in~\eq{se1}. The (I) red, (II) yellow, and (III) blue-thick curves correspond to the cases (I) without both the Re$(nkR)$-dependent localization enhancement $\Delta p$ and the periodic shift $\delta p$, (II) including $\Delta p$ and excluding $\delta p$, and (III) including both $\Delta p$ and $\delta p$. The robust agreement between the blue-thick curves and the numerical results in the upper panel and the discrepancies between the numerical results and the classical-reflection based curves corroborate the pivotal role of all semiclassical factors in predicting Re$(nkR)$-dependent Im$(nkR)$.}
    \label{fig:7}
\end{figure}

\section{Effects of semiclassical correction factors}\label{s7}
This section clarifies the impact of semiclassical factors considered in our semiclassical model for the Re$(nkR)$-dependent Im$(nkR)$ of group $G_3$. The thick curves in the upper panel of \fig{fig:7} represent the Re$(nkR)$-dependent Im$(nkR)$ obtained with Eq. (4) in the main text based on the semiclassical reflection Eq. (3) in the main text, while the lower panel shows the results obtained using the conventional classical Fresnel's reflection law, given by
     \begin{equation}
     \label{se1}
     \mathcal{Z}_\mathrm{classic}^\mathrm{TM}(\chi;n)=\left|\frac{n\cos\chi-\sqrt{1-n^2\sin^2\chi}}{n\cos\chi+\sqrt{1-n^2\sin^2\chi}}\right|^2.
     \end{equation}
Here, $\chi$ and $n=2$ represent the incident angle of the ray with respect to the inward-pointing normal vector of the boundary at which the ray is reflected and the refractive index, respectively. In each case, we compute Eq. (4) for three sub-cases. First, the red-thick curves in \fig{fig:7} show the case without both the Re$(nkR)$-dependent localization enhancement $\Delta p=0$ [cf. main text] and the periodic shift $\delta p=0$ [cf. main text]. Second, the yellow-thick curves in \fig{fig:7} show the cases of including $\Delta p$ and excluding $\delta p$. Lastly, the blue-thick curves in \fig{fig:7} represent the case where both $\Delta p$ and $\delta p$ are included.

Comparing all six cases in \fig{fig:7}, we conclude that considering all semiclassical factors is crucial to accurately reproduce the converging branches of the Re$(nkR)$-dependent decay rates Im$(nkR)$.
\end{widetext}
%\clearpage
%\newpage
\end{appendix}
\end{document}